 \documentstyle[prd,preprint,aps,epsf]{revtex}  
 \tightenlines
\begin{document}
 
\draft
\preprint{Submitted to {\em Physical Review D}}
\title{Photon splitting in strong magnetic fields: asymptotic
approximation formulae vs. accurate numerical results} 

\author{C. Wilke  and G. Wunner}
\address{Theoretische Physik I, Ruhr-Universit\"at Bochum,
         D-44780 Bochum, Germany}

%\date{Received } 
\maketitle

\begin{abstract}
We present the results of a numerical calculation of the photon
splitting rate below the electron-pair creation 
threshold ($\omega \le 2m$) in  
magnetic fields $B ~{> \atop \sim}~ B_{\rm cr} =   m^2 /e  = 
4.414 \times 10^{9}$ T. Our results confirm asymptotic approximations
derived in the
low-field ($B < B_{\rm cr}$) and high-field ($B \gg  B_{\rm cr}$) limit,
and allow interpolating between the two asymptotic regions. Our expression for 
the photon splitting rate is  a simplified version of a formula given 
by Mentzel et al. 
We also point out that, although the analytical formula is correct,
the splitting rates calculated there are wrong
due to an error in the numerical calculations.
\end{abstract}

\vskip 1.0 truecm


The exotic process of magnetic photon splitting, i.e.\ the decay of  a
photon into two   photons in the presence of a very strong
magnetic field, has recently attracted renewed attention, mainly
because of the great importance this process may have in the
interpretation of the spectra of cosmic $\gamma$-ray burst sources.
The basic formulae for magnetic photon splitting had already been 
derived in the seventies
\cite{Adler1,Bial_Birula,Adler2,Papanyan_Ritus,Stoneham}. In the first
approach the photon splitting effect was analyzed using
the Heisenberg-Euler effective Lagrangian. This method is
justified under the condition $\omega\ll m$. The result was the
$(\omega/m)^5(B/B_{\rm cr})^6$ dependency of the attenuation
coefficient in the weak field regime 
($B\ll B_{\rm cr}$)~\cite{Adler1,Bial_Birula}. Adler~\cite{Adler2} was
the first to solve the problem for arbitrary magnetic field strengths and 
photon energies below the pair 
creation threshold. He used the
gauge invariant proper-time method and presented numerical
results up to $B=B_{\rm cr}$ for the two cases $\omega=m$ and
$\omega\ll m$. Other gauge-invariant versions of the
splitting-amplitude were found later by
Stoneham~\cite{Stoneham}, Baier et al.~\cite{Baier2}, and recently,
using a path-integral approach, by Adler and Schubert~\cite{Adler3}. 
Nevertheless, numerical results, apart from those obtained by Adler, and
in particular for magnetic fields exceeding $B_{\rm cr}$, were not
available. Mentzel et al.~\cite {Mentzel1} therefore undertook a rederivation
of the photon splitting
rate using a configuration space representation for the electron propagator in
a strong   magnetic field with the aim to find an exact analytical
expression that would be suitable for numerical evaluation for
magnetic fields above $B_{\rm cr}$. The  representation chosen for   the
electron propagator, though not gauge-invariant, had the advantage
that the final expression for the photon splitting rate
did not contain any singularities for photon
energies below $\omega=2m$. Moreover, the sums over the Landau-states
that had to be calculated in this method were expected to converge
very rapidly for $B>B_{\rm cr}$.

The numerical results presented in~\cite {Mentzel1}
seemed to suggest  that the absolute values of the rates   were much bigger
than was previously assumed, but failed to reproduce the asymptotic 
low-frequency, low-field behavior that followed  from the earlier
investigatons. Consequently, these results were criticized by 
Adler \cite{ad96a}, who
suggested that the fixing of the gauge had led to calculation errors
that could not be detected in the final expression.

More recently, Baier et al. \cite {Baier1} determined 
the behavior of the photon splitting rate in the limit of extremely
high magnetic 
fields, $ B/B_{\rm cr} \gg 1$, 
and demonstrated that in this case the splitting rate
becomes independent of field strength. 
In agreement with these results are the recent results of Heyl and
Hernquist~\cite{Heyl} who reexamined the effective Lagrangian method
and were able to calculate the attenuation coefficient for arbitrary
magnetic field strengths and $\omega\ll m$.

In this note we will reexamine the
formula 
given by Mentzel et al. \cite {Mentzel1}, simplify it, evaluate it
numerically, and 
compare with the results of the asymptotic formulae  obtained by
Adler~\cite{Adler2}, Baier et al. \cite {Baier1}, and Heyl and 
Hernquist~\cite{Heyl}. This will allow  
to quantitatively assess the range of validity of the different
asymptotic results. We point out that
the numerical results of~\cite {Mentzel1} are in error because of an incorrect
sign in the code for evaluating 
the analytical formula for magnetic photon splitting.

Following the
paper by Baier et al. \cite {Baier1}  we shall 
concentrate on the process $ \perp\rightarrow\parallel\,\parallel$, 
where a photon with the polarization vector
parallel to  the plane formed by the  $\vec k$  vector and the
magnetic field axis
decays into two photons with polarization vectors perpendicular to that 
plane. This is expected to be the dominating decay channel as soon as 
dispersive effects become important. Note that these designations for the
polarizations are exactly opposite to those chosen by Adler. Furthermore, 
we  confine ourselves to photon energies below the pair creation
threshold $\omega < 2 m$, above which pair creation is expected to  dominate
photon splitting.

The complexity of the general expressions for the photon splitting rate
in magnetic fields has   always been an impediment to quantitative
studies of the 
importance of the photon splitting process in actual astrophysical model
calculations. The availability of appropriate approximation formulae, and 
a knowledge of their respective ranges of applicability, is therefore 
a prerequisite for such investigations \cite{Baring}. On the low-frequency,
weak-field  ($B ~{< \atop \sim}~ B_{\rm cr}$) side, the attenuation
coefficient can be written (in the notation of Stoneham \cite{Stoneham}) 
\begin{equation}
\ell^{-1}(\perp\to \| ~ \|) = {{\alpha^3 {m}} \over {60 \pi^2}} 
    \left ( B \over {B_{\rm cr}} \right)^6  \left ( \omega \over {m} 
 \right)^5 
     \times \left(M_1(B) \right)^2 \;,
\label{perturb}
\end{equation}      
where the scattering amplitude $\left(M_1(B) \right)$ is given by eq. 41
of ref. \cite{Stoneham}. Adler~\cite{Adler2}
demonstrated that   
for field strenghts up to $\sim 0.1~B_{\rm cr}$ the coefficient
is excellently reproduced by that equation, with $M_1$ replaced with its
$B = 0$ value, $M_1(0) = 26/315$.

On the very high field side, $B \gg B_{\rm cr}$ Baier et al. \cite{Baier1} 
have recently shown that, for given
energy below $2m$, the attenuation coefficent tends towards a
constant, magnetic field  
independent value, viz.
\begin{equation}
\ell^{-1}(\perp\to \| ~ \|) = \int \frac {1}{32\pi}|T_{B \gg B_{\rm
    cr}}|^2\frac {{\rm d}\omega'}{\omega^2},
\label{formel-baier1}
\end{equation}     
where $T_{B \gg B_{\rm cr}}$ is the high-field approximation of the
splitting amplitude: 
\begin{eqnarray}
T_{B \gg B_{\rm cr}} &=& \frac {2(4\pi\alpha)^{3/2}}{\pi^2}
     \Big[\frac {\omega'm^2}{\omega''\sqrt{4m^2-\omega''^2}}
                  \arctan\left(\frac
                  {\omega''}{\sqrt{4m^2-\omega''^2}}\right)
                  \nonumber\\
&& \qquad
      + \frac {\omega''m^2}{\omega'\sqrt{4m^2-\omega'^2}}
                  \arctan\left(\frac
                  {\omega'}{\sqrt{4m^2-\omega'^2}}\right)
      -\frac {1}{4} \omega \Big]\;.
\label{formel-baier2}
\end{eqnarray}   

On the basis of an accurate numerical evaluation of the general expression for 
the photon splitting  rate we will now  determine the range of field
strengths and frequencies from where on the high-field asymptotic behavior is 
valid, and provide data for the range of intermediate field strengths. 

The general expression we
evaluate is the one derived in \cite {Mentzel1}. Two
important remarks  
on this work are in order. The first
is that by reason 
of a simple error in sign in the computer code (the quantity $K_2$ defined 
in eq. (A12) in the Appendix of ref. \cite{Mentzel1} contained a plus sign
in the code instead of a minus sign in the first term on the
right-hand side) all the  
{\em numerical} results are wrong. This error escaped the numerical
checks of the 
polarization selection rules. Speculations that gauge invariance problems
caused the deviations from previous results are thus found to be
insubstantial.  
The second
remark is that the {\em analytical} expression obtained by  Mentzel et
al. \cite {Mentzel1} is  
{\em correct}. 
The low-frequency, weak-field limit (\ref{perturb}) can be derived
from it analytically, 
and the results of the numerical evaluation (corrected for the error in sign)
are in agreement with previous results in all ranges of parameters where 
a comparison is possible. We note  that by a suitable rearrangement  
of terms the numerical
evaluation of the expression of~\cite {Mentzel1} could
be speeded up by 
two orders of magnitudes. The main step in these optimizations is the
summing up of the different polarization
states. We performed this sum using the computer algebra program
Maple. Once  the summation has been carried out, the remaining integration 
can be
reduced to 5 integrals that can be evaluated independendly from the
rest of the formula.

An important point with regard to  the
accuracy of the results 
is the inclusion of sufficiently many  Landau states in carrying out the
sums over the intermediate states. In particular for magnetic fields 
in excess of $B_{\rm cr}$ these sums are found to converge very rapidly.

In Fig.~1 we present, for the three frequencies $\omega/m =$ 0.1, 1.0, 1.99,  
the results of the numerical evaluation (crosses) of the
general expression for the  magnetic photon
splitting  attenuation coefficient for the decay channel  
$\perp\to \| ~ \|$ as a function of the field strength 
in comparison with the predictions of the 
asymptotic formulae. The steep solid line on the left corresponds to the 
$\omega^5 \, B^6$ dependence of the low-field formula
eq. (\ref{perturb}) (with $M_1(0)$ 
inserted for $M_1(B))$, while the dotted horizontal line corresponds to the
high-field limit (\ref{formel-baier1}). The figure shows how, for all three
frequencies, the numerical results approach the low-field limit as $B$
decreases, 
and that the numerical values tend towards  the high-field limit as
$B$ increases. The results are in good agreement with those of
Heyl and Hernquist~\cite{Heyl}. We find that their formulae have a very
high accuracy for photon energies up to $\omega=0.5m$.
The high-field limit of Baier et al.~\cite{Baier2} is
reached, to within a few per cent, already for magnetic field strengths
$B/B_{\rm cr}$  around  30, independent of the photon-energy.

We now proceed to results for the differential splitting rate.   
This quantity provides the information
on the probability with which a certain partitioning of the
energy of the decaying photon on the outgoing photons is
assumed. Therefore it is reasonable to normalize this quantity to unity.
Note that in this respect our procedure differs from that of Baier et
al. \cite{Baier1}, who used a magnetic-field dependent normalization
constant, which blurs the comparison of the results at different field
strengths.  
 
In Fig.~2,  for  the   energies $\omega/m=0.1$, $\omega/m=1.5$ and
$\omega/m=1.99$ and magnetic field strengths  $B= B_{\rm cr}, ~5~B_{\rm
  cr}$ and $100~B_{\rm cr}$
 numerical results are shown for the differential
splitting rate (normalized to unity when integrated over the abscissa)
as a function of  $\omega^\prime/\omega$. For $B=100~B_{\rm cr}$, as
expected, the shape of the differential
splitting rate is exactly identical, for all three photon energies, 
to the one determined from  Baier's asymptotic expression (solid
curves in Fig.~2).  
This  agreement, however, is seen to be present even for the smaller magnetic
field strengths of $B = 5~B_{\rm cr}$ and   $B= B_{\rm cr}$, and it is
only  for  
photon energies approaching the pair creation threshold  in the latter 
case  that sizeable deviations from
the  asymptotic behavior appear. We also see from Fig.~2 that    
the plateau-forming effect described by Baier et al.~\cite {Baier1} occurs
already below $B=5~B_{\rm cr}$. Thus we have the result that the high-field
asymptotic formulae represent a good approximation even at field
strengths where  
they  were not a priori expected to apply. This opens the possibility of
carrying out astrophysical model calculations using the asymptotic
high-field formulae for $B ~{> \atop \sim}~ 5 B_{\rm cr}$, field
strengths that are supposed to be present in  soft $\gamma$ repeaters
\cite{Baring,Duncan}.

In this note we have presented numerically accurate values of 
attenuation coefficients for photon
splitting in   strong magnetic fields, and have used these results to
determine the 
ranges of parameters in which    asymptotic approximation
formulae can be applied in the place of the full complicated expression
for the photon splitting rate.   The  results should stimulate, in high-energy 
astrophysics, new quantitative studies of the role of the exotic
process of photon  
splitting in  the formation of high-energy spectra  of   strongly
magnetized cosmic  
objects.
\acknowledgements  
We thank Dr. Michael Kachelrie{\ss} for stimulating discussions.

\begin{figure}
 \centerline{
 \epsfxsize=3in
 \epsffile{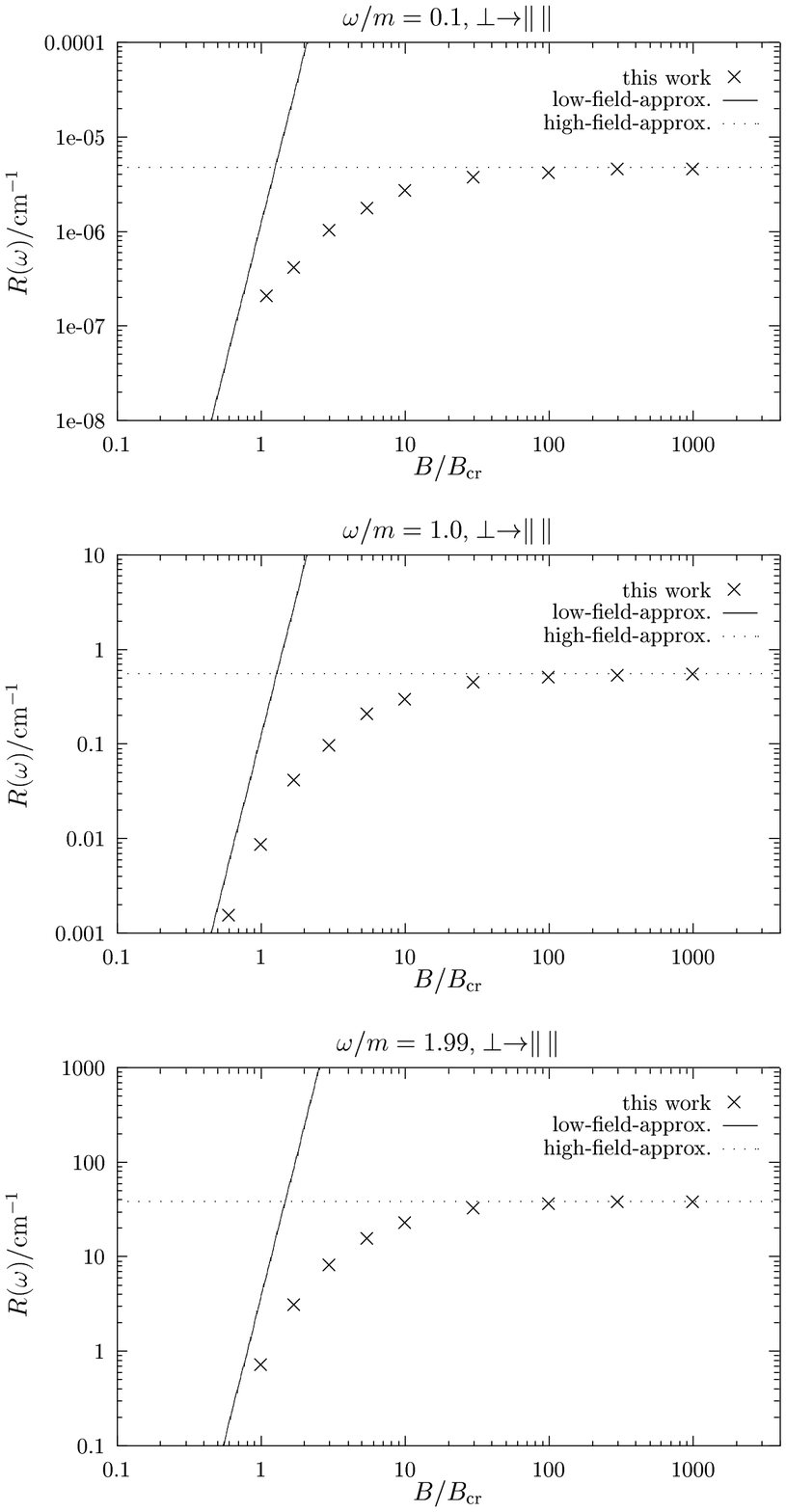}
 }
 \bigskip\noindent
\caption{Total attenuation coefficient  for the splitting
channel $\perp\rightarrow\parallel\,\parallel$ as a  function  of
field strength 
for  the three photon frequencies $\omega/m= $ 0.1, 1.0, and 1.99.
Crosses denote the results of the present calculations, the weak-field 
result  is represented by the solid straight line, the high-field
approximation by the horizontal dotted line.}
\end{figure}

\begin{figure}
 \centerline{
 \epsfxsize=3in
 \epsffile{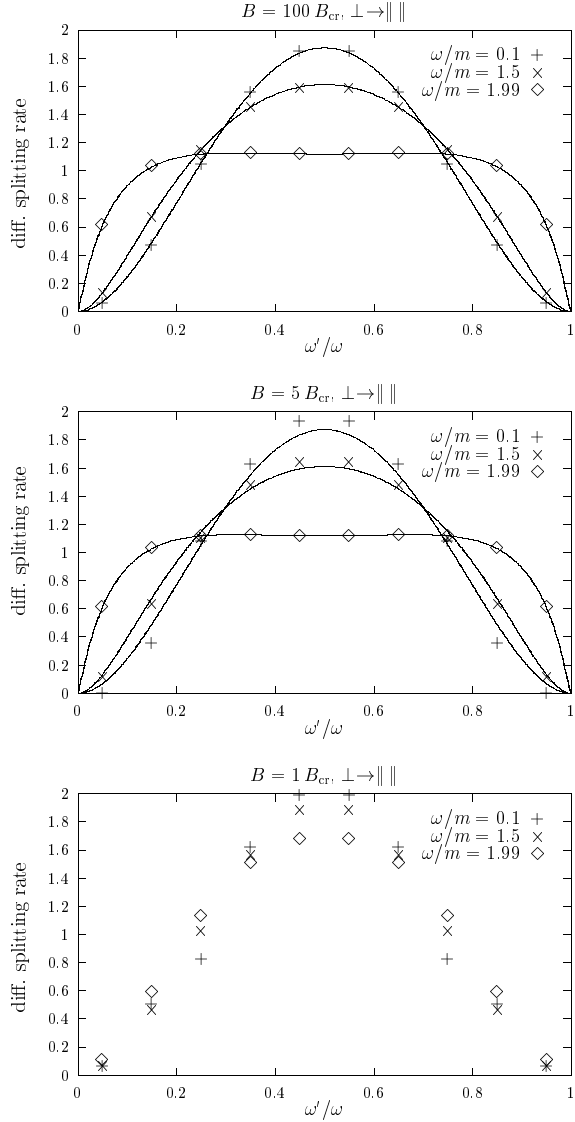}
 }
 \bigskip\noindent
\caption{Differential splitting rate (normalized to
unity when integrated over the abscissa) for the splitting channel
$\perp\rightarrow\parallel\,\parallel$ and three frequencies
$\omega$ of the incoming photon as a function of the 
ratio of the frequency $\omega^\prime$ of one of the outgoing photons
and $\omega$,    
for the magnetic field strengths (from top to bottom)
$B/B_{\rm cr}$ = 100, 5, and 1. The solid curves represent the 
results of the asymptotic high-field approximation by Baier et al. 
This
approximation is seen to be valid even at magnetic field strengths
as low as $B \sim 5 B_{\rm cr}$.}
\end{figure}
\end {document}